# Mathematical Formulae for the Vibration Frequencies of Rubber Wiper on Windshield

Ying-Ji Hong [1] and Tsai-Jung Chen [2]

[1] Department of Mathematics, National Cheng-Kung University, Taiwan

[2] Department of Vehicle Engineering, National Pingtung University of Science and Technology, Taiwan

Corresponding author:
Ying-Ji Hong, Department of Mathematics, National Cheng-Kung University, No.1, University Road, Tainan City 70101, Taiwan
Email: yjhong@mail.ncku.edu.tw

**Abstract**

Automotive engineers want to reduce the noise generated by the vibrations of rubber wiper on the windshield of an automobile. To understand the vibrations of wiper noise, certain spring-mass models were presented by some specialists, over the past few years, to simulate the vibrations of rubber wiper on windshield.

In this article, we will give precise mathematical formulae for the vibration frequencies of rubber wiper on windshield. Comparison of our model predictions with experimental data confirms the accuracy of our mathematical formulae for the vibration frequencies of wiper on windshield. In fact, our model predictions are in almost perfect agreement with experimental data.

These mathematical formulae for the vibration frequencies of rubber wiper on windshield are derived from our analysis of a 3-dimensional elastic model with specific boundary conditions. These specific boundary conditions are set up due to mechanical and mathematical consideration. Our mathematical formulae can be used to test the quality of wiper design.

## 1. Introduction

Automotive engineers want to reduce the noise generated by the vibrations of rubber wiper on the windshield of an automobile. To understand the vibrations of wiper noise, certain spring-mass models were presented by some specialists, over the past few years, to simulate the vibrations of rubber wiper on windshield (Stein et al., 2008; Sugita et al., 2012; Reddyhoff et al., 2015; Lancioni et al., 2016; Unno et al, 2017).

In this article, we will analyze the vibration frequencies of rubber wiper on windshield through a 3-dimensional physical model. Our analysis leads to mathematical formulae for the vibration frequencies of rubber wiper on windshield.

Let $\rho$ denote the *density* of the rubber wiper. Let $l$ denote the length of the rubber wiper. Our analysis shows that there exist two classes of vibration frequencies of rubber wiper. For the vibration frequencies of Class I, we predict that the vibration frequencies of rubber wiper will locate around

$$\sqrt{\frac{\lambda + 2\mu}{\rho}} \cdot \frac{n}{2l} \text{ Hz} \quad \text{(Class I)} \tag{1}$$

where $n$ is a nonnegative integer. For the vibration frequencies of Class II, we predict that the vibration frequencies of rubber wiper will locate around

$$\sqrt{\frac{\mu}{\rho}} \cdot \frac{n}{2l} \text{ Hz} \quad \text{(Class II)}. \tag{2}$$

Here $\lambda$ and $\mu$ are Lame coefficients. Lame coefficients are material constants of the rubber wiper. These material constants are related to the Young's modulus $E$ and the Poisson's ratio $\sigma$ as follows (Chaichian et al., 2012):

$$\lambda = \frac{\sigma \cdot E}{(1+\sigma) \cdot (1-2\sigma)} \quad \text{and} \quad \mu = \frac{E}{2 \cdot (1+\sigma)}. \tag{3}$$

Experimental data highly supports the effectiveness of our model. In fact, our model predictions are in almost perfect agreement with experimental data.

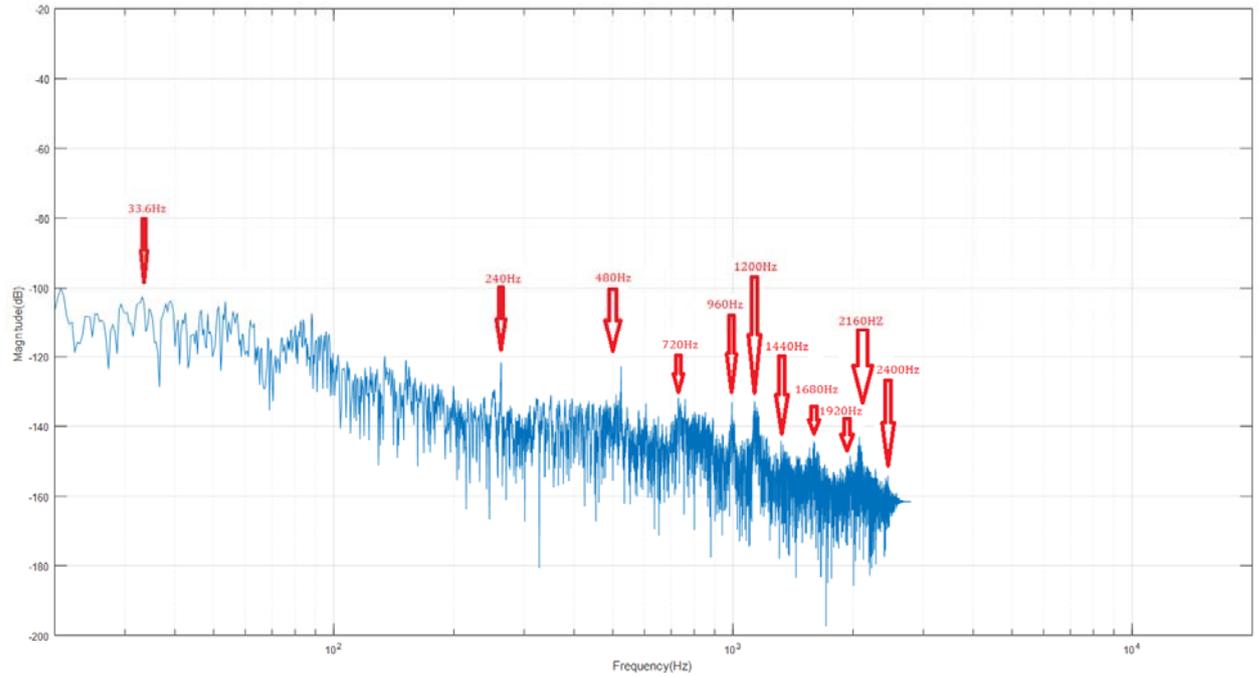

**Figure 1.** Vibration frequencies of rubber wiper before reversal.

For the material parameters of our rubber wiper, our calculation shows that the vibration frequencies of Class I will locate around 240Hz, 480Hz, 720Hz, 960Hz, 1200Hz, 1440Hz, 1680Hz, 1920Hz, 2160Hz, 2400Hz, etc. Our calculation shows that the vibration frequencies of Class II will locate around $n \cdot (33.6Hz)$. The predicted peaks of frequencies are indicated on the FFT (Fast Fourier Transform) diagram Figure 1 of our experimental data. In this FFT diagram, magnitude of FFT for frequencies below 25,600Hz can be trusted. Details of our calculation will be explained in Section 3.

## 2. Physical model for the vibrations of rubber wiper on windshield

In this section, we discuss our 3-dimensional physical model for the vibrations of rubber wiper on windshield. We will introduce the Lame system of partial differential equations with specific boundary conditions on a 3-dimensional rectangular solid plate $P$ in Subsection 2.1.

There are two wave equations with different wave speeds associated with the Lame system. Assume that the smooth solutions of the wave equation with *higher* wave speed are collected in Class I. Assume that the smooth solutions of the wave equation with *lower* wave speed are collected in Class II. It is known that the solutions of Class I and the solutions of Class II can be added to generate all smooth vector-valued solutions of the Lame system (Hetnarski et al., 2004; Hong et al., 2020).

On the other hand, it is known that all the smooth solutions of a one-dimensional wave equation can be expressed as the infinite sums of special solutions for this wave equation found through the Fourier method (Strauss, 2008; Serov, 2017).

Thus, in Subsection 2.2 and Subsection 2.3, we will respectively construct special solutions of Class I and Class II for the Lame system with specific boundary conditions on the rectangular solid plate $P$. The *vibration frequencies* of these special solutions of Class I and Class II will be shown explicitly.

In Subsection 2.4, we will discuss the physical significance of some important parameters appearing in our construction of the special solutions of Class I and Class II. We will then explain how to arrive at the mathematical formulae (1) and (2) through physical consideration.

## 2.1. Elastic plate model

When a rubber wiper blade moves on the windshield of an automobile, there are 4 different forces acting on it: pressure on the wiper blade, support force from the windshield, drag force, and the (kinetic) friction force. Figure 2 is a 2-dimensional force diagram for the rubber wiper blade.

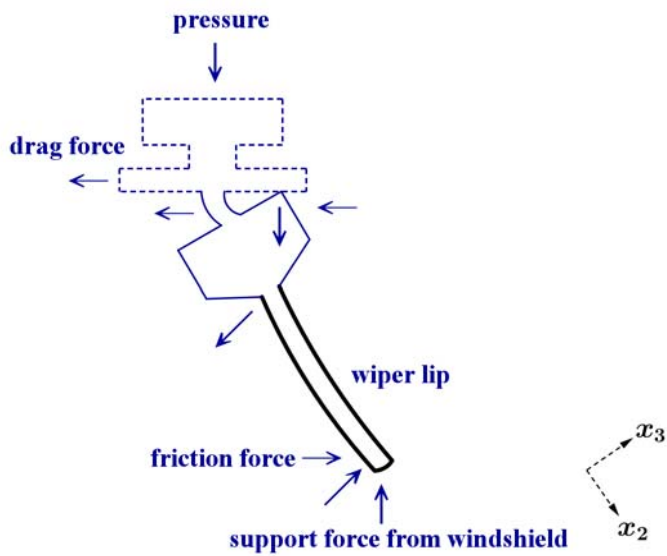

**Figure 2.** Two-dimensional force diagram for the rubber wiper blade on windshield.

To understand the vibrations of rubber wiper blade, it is natural to consider the vibrations of wiper lip. Thus we consider, for simplicity, a 3-dimensional rectangular solid plate $P$ with length $l$ m, width $w$ m, and thickness $h$ m. See Figure 3.

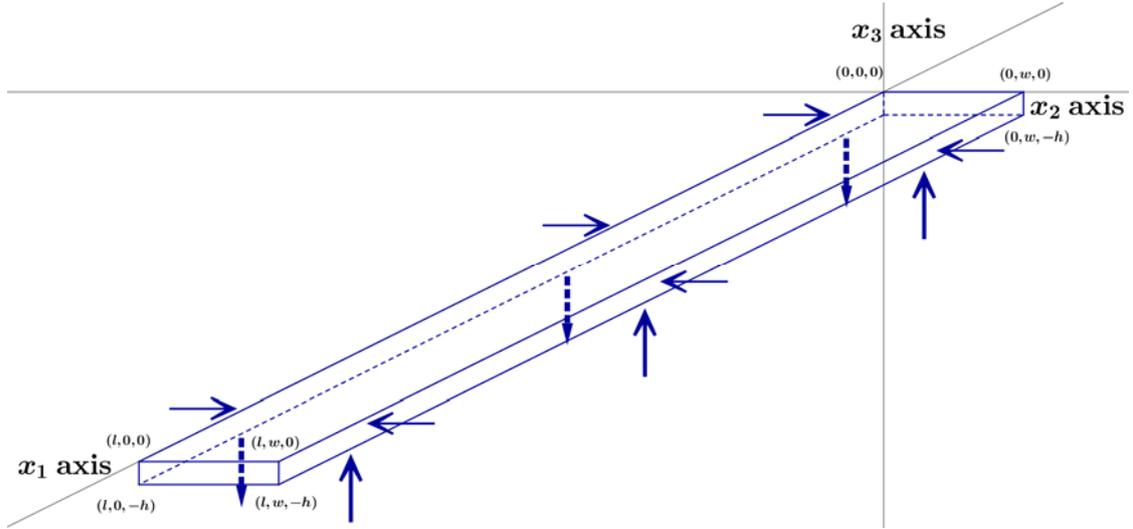

**Figure 3.** Rectangular plate with length $l$ m, width $w$ m, and thickness $h$ m.

On this hyper-elastic plate, all the forces acting on wiper lip will be considered as distribution of traction/stress force. In fact, it is known that the "friction force" should be considered, from a physical point of view, as "distribution of intermolecular forces" (Yang et al., 2008). Thus the traction/stress distribution on the boundary of our 3-dimensional elastic model could be discontinuous.

Let $\mathbf{Z}_{bottom}$, $\mathbf{Z}_{ll}$, and $\mathbf{Z}_{rl}$ denote respectively the bottom, left lateral, and right lateral rectangles of the boundary surface of $\mathbf{P}$ defined as follows.

$$\mathbf{Z}_{bottom} = \{(x_1, x_2, -h) : 0 \leq x_1 \leq l \text{ and } 0 \leq x_2 \leq w\}. \tag{4}$$

$$\mathbf{Z}_{ll} = \{(x_1, 0, x_3) : 0 \leq x_1 \leq l \text{ and } -h \leq x_3 \leq 0\}. \tag{5}$$

$$\mathbf{Z}_{rl} = \{(x_1, w, x_3) : 0 \leq x_1 \leq l \text{ and } -h \leq x_3 \leq 0\}. \tag{6}$$

As shown in Figure 2 and Figure 3, the *traction force* acts on the rectangles $\mathbf{Z}_{bottom}$, $\mathbf{Z}_{ll}$, and $\mathbf{Z}_{rl}$ of the boundary surface of $\mathbf{P}$.

Let $\mathbf{Z}_{top}$, $\mathbf{Z}_{anterior}$, and $\mathbf{Z}_{posterior}$ denote respectively the top, anterior, and posterior rectangles of the boundary surface of $\mathbf{P}$ defined as follows.

$$\mathbf{Z}_{top} = \{(x_1, x_2, 0) : 0 \leq x_1 \leq l \text{ and } 0 \leq x_2 \leq w\}. \tag{7}$$

$$\mathbf{Z}_{anterior} = \{(l, x_2, x_3) : 0 \leq x_2 \leq w \text{ and } -h \leq x_3 \leq 0\}. \tag{8}$$

$$\mathbf{Z}_{posterior} = \{(0, x_2, x_3) : 0 \leq x_2 \leq w \text{ and } -h \leq x_3 \leq 0\}. \tag{9}$$

As shown in Figure 3, the traction force does *not* act on the rectangles $\mathbf{Z}_{top}$, $\mathbf{Z}_{anterior}$, and $\mathbf{Z}_{posterior}$ of the boundary surface of $\mathbf{P}$.

To determine the vibrations of rubber wiper blade, we consider the vector-valued *displacement* function

$$\boldsymbol{u}(t,\boldsymbol{x}) = \left(u_1(t,\boldsymbol{x}), u_2(t,\boldsymbol{x}), u_3(t,\boldsymbol{x})\right) \quad \text{with} \quad \boldsymbol{x} = (x_1, x_2, x_3) \in \boldsymbol{P}. \tag{10}$$

The dynamics of rubber wiper blade is *completely determined* by the vector-valued displacement function $\boldsymbol{u}(t,\boldsymbol{x})$. This displacement function $\boldsymbol{u}(t,\boldsymbol{x})$ must satisfy the following *vector-valued* Lame system (of 3 interdependent partial differential equations)

$$\frac{\partial^2 \boldsymbol{u}}{\partial t^2} = \frac{(\lambda+\mu)}{\rho} \cdot \nabla\left(\frac{\partial u_1}{\partial x_1} + \frac{\partial u_2}{\partial x_2} + \frac{\partial u_3}{\partial x_3}\right) + \frac{\mu}{\rho} \cdot \left(\frac{\partial^2 \boldsymbol{u}}{\partial x_1^2} + \frac{\partial^2 \boldsymbol{u}}{\partial x_2^2} + \frac{\partial^2 \boldsymbol{u}}{\partial x_3^2}\right)$$

$$= \frac{(\lambda+\mu)}{\rho} \cdot \nabla(\operatorname{div} u) + \frac{\mu}{\rho} \cdot \nabla^2 \boldsymbol{u} \tag{11}$$

(Frankel, 2011; Chaichian et al., 2012). Since the distribution of *shear* stress on the boundary surface of the solid plate $\boldsymbol{P}$ is not continuous, we will only require, due to mathematical consideration, the following boundary conditions (11), (12), and (13).

$$\lambda \cdot \left(\frac{\partial u_1}{\partial x_1} + \frac{\partial u_2}{\partial x_2} + \frac{\partial u_3}{\partial x_3}\right) + 2\mu \cdot \frac{\partial u_3}{\partial x_3} = 0 \quad \text{when} \quad x_3 = 0. \tag{12}$$

$$\lambda \cdot \left(\frac{\partial u_1}{\partial x_1} + \frac{\partial u_2}{\partial x_2} + \frac{\partial u_3}{\partial x_3}\right) + 2\mu \cdot \frac{\partial u_1}{\partial x_1} = 0 \quad \text{when} \quad x_1 = l. \tag{13}$$

$$\lambda \cdot \left(\frac{\partial u_1}{\partial x_1} + \frac{\partial u_2}{\partial x_2} + \frac{\partial u_3}{\partial x_3}\right) + 2\mu \cdot \frac{\partial u_1}{\partial x_1} = 0 \quad \text{when} \quad x_1 = 0. \tag{14}$$

## 2.2. Solutions of Class I

Assume that $\boldsymbol{u}(t,\boldsymbol{x}) = \left(u_1(t,\boldsymbol{x}), u_2(t,\boldsymbol{x}), u_3(t,\boldsymbol{x})\right)$ is the *displacement* function. We define the component functions of $\boldsymbol{u}(t,\boldsymbol{x})$ as follows.

$$u_1(t,\boldsymbol{x}) = \frac{n \cdot \pi}{l} \cdot \sin(\varphi + k \cdot t) \cdot \cos\left(\frac{n \cdot \pi \cdot x_1}{l}\right) \cdot \sin(a + \alpha \cdot x_2) \cdot \sin(\beta \cdot x_3). \tag{15}$$

$$u_2(t,\boldsymbol{x}) = \alpha \cdot \sin(\varphi + k \cdot t) \cdot \sin\left(\frac{n \cdot \pi \cdot x_1}{l}\right) \cdot \cos(a + \alpha \cdot x_2) \cdot \sin(\beta \cdot x_3). \tag{16}$$

$$u_3(t,\boldsymbol{x}) = \beta \cdot \sin(\varphi + k \cdot t) \cdot \sin\left(\frac{n \cdot \pi \cdot x_1}{l}\right) \cdot \sin(a + \alpha \cdot x_2) \cdot \cos(\beta \cdot x_3). \tag{17}$$

Here the set $(n, \alpha, \beta, k)$ of parameters must satisfy

$$k^2 = \frac{\lambda + 2\mu}{\rho} \cdot \left(\frac{n^2 \cdot \pi^2}{l^2} + \alpha^2 + \beta^2\right) \tag{18}$$

in which $n$ is an *integer*. It can be checked readily that the vector-valued function $\boldsymbol{u}(t,\boldsymbol{x})$ defined by (15), (16), and (17) is a solution for (11), (12), (13), and (14). The frequency of this solution $u(t,x)$ is

$$v = \frac{|k|}{2\pi} = \sqrt{\frac{\lambda+2\mu}{\rho}} \cdot \sqrt{\left(\frac{n^2 \cdot \pi^2}{4 \cdot \pi^2 \cdot l^2} + \frac{\alpha^2}{4 \cdot \pi^2} + \frac{\beta^2}{4 \cdot \pi^2}\right)}. \tag{19}$$

When $|\alpha|$ and $|\beta|$ are relatively small, we have

$$v \approx \sqrt{\frac{\lambda+2\mu}{\rho}} \cdot \frac{|n|}{2l}. \tag{20}$$

In general, we may consider $u(t,x) = (u_1(t,x), u_2(t,x), u_3(t,x))$ defined as follows.

$$u_1(t,x) = \frac{n \cdot \pi}{l} \cdot \sin(\varphi + k \cdot t) \cdot \cos\left(\frac{n \cdot \pi \cdot x_1}{l}\right) \cdot f(a + \alpha \cdot x_2) \cdot g(\beta \cdot x_3). \tag{21}$$

$$u_2(t,x) = \alpha \cdot \sin(\varphi + k \cdot t) \cdot \sin\left(\frac{n \cdot \pi \cdot x_1}{l}\right) \cdot f'(a + \alpha \cdot x_2) \cdot g(\beta \cdot x_3). \tag{22}$$

$$u_3(t,x) = \beta \cdot \sin(\varphi + k \cdot t) \cdot \sin\left(\frac{n \cdot \pi \cdot x_1}{l}\right) \cdot f(a + \alpha \cdot x_2) \cdot g'(\beta \cdot x_3). \tag{23}$$

Here $n$ must be an *integer*. We discuss the choice of $f$ and $g$ in what follows.

• **Case I**: $f = \sin$ and $g = \sin$. In this case we arrive at the displacement function $u(t,x)$ defined by (15), (16), and (17), with the set $(n, \alpha, \beta, k)$ of parameters satisfying (18).

• **Case II**: $f = \sin$ and $g = \sinh$. In this case the set $(n, \alpha, \beta, k)$ of parameters must satisfy

$$k^2 = \frac{\lambda+2\mu}{\rho} \cdot \left(\frac{n^2 \cdot \pi^2}{l^2} + \alpha^2 - \beta^2\right) \geq 0. \tag{24}$$

With (24) being satisfied, it can be checked readily that the vector-valued function $u(t,x)$ defined by (21), (22), and (23) is a solution for (11), (12), (13), and (14). The frequency of this solution is

$$v = \frac{|k|}{2\pi} = \sqrt{\frac{\lambda+2\mu}{\rho}} \cdot \sqrt{\left(\frac{n^2 \cdot \pi^2}{4 \cdot \pi^2 \cdot l^2} + \frac{\alpha^2}{4 \cdot \pi^2} - \frac{\beta^2}{4 \cdot \pi^2}\right)}. \tag{25}$$

• **Case III**: $f = \sinh$ or $f = \cosh$ with $g = \sin$. In this case the set $(n, \alpha, \beta, k)$ of parameters must satisfy

$$k^2 = \frac{\lambda+2\mu}{\rho} \cdot \left(\frac{n^2 \cdot \pi^2}{l^2} - \alpha^2 + \beta^2\right) \geq 0. \tag{26}$$

With (26) being satisfied, it can be checked readily that the vector-valued function

$u(t, x)$ defined by (21), (22), and (23) is a solution for (11), (12), (13), and (14). The frequency of this solution is

$$v = \frac{|k|}{2\pi} = \sqrt{\frac{\lambda + 2\mu}{\rho}} \cdot \sqrt{\left(\frac{n^2 \cdot \pi^2}{4 \cdot \pi^2 \cdot l^2} - \frac{\alpha^2}{4 \cdot \pi^2} + \frac{\beta^2}{4 \cdot \pi^2}\right)}. \tag{27}$$

● **Case IV**: $f = \sinh$ or $f = \cosh$ with $g = \sinh$. In this case the set $(n, \alpha, \beta, k)$ of parameters must satisfy

$$k^2 = \frac{\lambda + 2\mu}{\rho} \cdot \left(\frac{n^2 \cdot \pi^2}{l^2} - \alpha^2 - \beta^2\right) \geq 0. \tag{28}$$

With (28) being satisfied, it can be checked readily that the vector-valued function $u(t, x)$ defined by (21), (22), and (23) is a solution for (11), (12), (13), and (14). The frequency of this solution is

$$v = \frac{|k|}{2\pi} = \sqrt{\frac{\lambda + 2\mu}{\rho}} \cdot \sqrt{\left(\frac{n^2 \cdot \pi^2}{4 \cdot \pi^2 \cdot l^2} - \frac{\alpha^2}{4 \cdot \pi^2} - \frac{\beta^2}{4 \cdot \pi^2}\right)}. \tag{29}$$

When $|\alpha|$ and $|\beta|$ are relatively small, we have the following estimate for the vibration frequency of $u(t, x)$:

$$v \approx \sqrt{\frac{\lambda + 2\mu}{\rho}} \cdot \frac{|n|}{2l} \tag{30}$$

for all the cases discussed above.

**Remark 1.** When the requirement $k^2 \geq 0$ is not satisfied in any of the last three cases,

$$k^2 = \frac{\lambda + 2\mu}{\rho} \cdot \left(\frac{n^2 \cdot \pi^2}{l^2} \pm \alpha^2 \pm \beta^2\right) \prec 0, \tag{31}$$

so that $k$ is *purely imaginary*, the displacement function $u(t, x)$ defined by (21), (22), and (23) is still a solution for (11), (12), (13), and (14). However, the corresponding displacement function $u(t, x)$ would grow exponentially and could lead to *failure* of rubber wiper.

## 2.3. Solutions of Class II

Assume that $u(t, x) = (u_1(t, x), u_2(t, x), u_3(t, x))$ is the *displacement* function. We define the component functions of $u(t, x)$ as follows.

$$u_1(t, x) = \gamma_1 \cdot \sin(\varphi + k \cdot t) \cdot \cos\left(\frac{n \cdot \pi \cdot x_1}{l}\right) \cdot \sin(a + \alpha \cdot x_2) \cdot \sin(\beta \cdot x_3). \tag{32}$$

$$u_2(t,\boldsymbol{x}) = \gamma_2 \cdot \sin(\varphi + k \cdot t) \cdot \sin\left(\frac{n \cdot \pi \cdot x_1}{l}\right) \cdot \cos(a + \alpha \cdot x_2) \cdot \sin(\beta \cdot x_3). \tag{33}$$

$$u_3(t,\boldsymbol{x}) = \gamma_3 \cdot \sin(\varphi + k \cdot t) \cdot \sin\left(\frac{n \cdot \pi \cdot x_1}{l}\right) \cdot \sin(a + \alpha \cdot x_2) \cdot \cos(\beta \cdot x_3). \tag{34}$$

Here the sets $(\gamma_1, \gamma_2, \gamma_3)$ and $(n, \alpha, \beta, k)$ of parameters must satisfy respectively

$$\gamma_1 \cdot \frac{n \cdot \pi}{l} + \gamma_2 \cdot \alpha + \gamma_3 \cdot \beta = 0 \tag{35}$$

and

$$k^2 = \frac{\mu}{\rho} \cdot \left(\frac{n^2 \cdot \pi^2}{l^2} + \alpha^2 + \beta^2\right) \tag{36}$$

in which $n$ is an *integer*. It can be checked readily that the vector-valued function $\boldsymbol{u}(t,\boldsymbol{x})$ defined by (32), (33), and (34) is a solution for (11), (12), (13), and (14). The frequency of this solution is

$$\nu = \frac{|k|}{2\pi} = \sqrt{\frac{\mu}{\rho}} \cdot \sqrt{\left(\frac{n^2 \cdot \pi^2}{4 \cdot \pi^2 \cdot l^2} + \frac{\alpha^2}{4 \cdot \pi^2} + \frac{\beta^2}{4 \cdot \pi^2}\right)}. \tag{37}$$

In general, we may consider $\boldsymbol{u}(t,\boldsymbol{x}) = (u_1(t,\boldsymbol{x}), u_2(t,\boldsymbol{x}), u_3(t,\boldsymbol{x}))$ defined as follows.

$$u_1(t,\boldsymbol{x}) = \gamma_1 \cdot \sin(\varphi + k \cdot t) \cdot \cos\left(\frac{n \cdot \pi \cdot x_1}{l}\right) \cdot f(a + \alpha \cdot x_2) \cdot g(\beta \cdot x_3). \tag{38}$$

$$u_2(t,\boldsymbol{x}) = \gamma_2 \cdot \sin(\varphi + k \cdot t) \cdot \sin\left(\frac{n \cdot \pi \cdot x_1}{l}\right) \cdot f'(a + \alpha \cdot x_2) \cdot g(\beta \cdot x_3). \tag{39}$$

$$u_3(t,\boldsymbol{x}) = \gamma_3 \cdot \sin(\varphi + k \cdot t) \cdot \sin\left(\frac{n \cdot \pi \cdot x_1}{l}\right) \cdot f(a + \alpha \cdot x_2) \cdot g'(\beta \cdot x_3). \tag{40}$$

Here $n$ must be an *integer*. We discuss the choice of $f$ and $g$ in what follows.

● **Case I**: $f = \sin$ and $g = \sin$. In this case we arrive at the displacement function $\boldsymbol{u}(t,\boldsymbol{x})$ defined by (32), (33), and (34), with the sets $(\gamma_1, \gamma_2, \gamma_3)$ and $(n, \alpha, \beta, k)$ of parameters respectively satisfying (35) and (36).

● **Case II**: $f = \sin$ and $g = \sinh$. In this case the sets $(\gamma_1, \gamma_2, \gamma_3)$ and $(n, \alpha, \beta, k)$ of parameters must respectively satisfy

$$\gamma_1 \cdot \frac{n \cdot \pi}{l} + \gamma_2 \cdot \alpha - \gamma_3 \cdot \beta = 0 \text{ and } k^2 = \frac{\mu}{\rho} \cdot \left(\frac{n^2 \cdot \pi^2}{l^2} + \alpha^2 - \beta^2\right) \geq 0. \tag{41}$$

With (41) being satisfied, it can be checked readily that the vector-valued function $\boldsymbol{u}(t,\boldsymbol{x})$ defined by (38), (39), and (40) is a solution for (11), (12), (13), and (14). The frequency of this solution is

$$v = \frac{|k|}{2\pi} = \sqrt{\frac{\mu}{\rho}} \cdot \sqrt{\left(\frac{n^2 \cdot \pi^2}{4 \cdot \pi^2 \cdot l^2} + \frac{\alpha^2}{4 \cdot \pi^2} - \frac{\beta^2}{4 \cdot \pi^2}\right)}. \tag{42}$$

● **Case III**: $f = \sinh$ or $f = \cosh$ with $g = \sin$. In this case the sets $(\gamma_1, \gamma_2, \gamma_3)$ and $(n, \alpha, \beta, k)$ of parameters must respectively satisfy

$$\gamma_1 \cdot \frac{n \cdot \pi}{l} - \gamma_2 \cdot \alpha + \gamma_3 \cdot \beta = 0 \quad \text{and} \quad k^2 = \frac{\mu}{\rho} \cdot \left(\frac{n^2 \cdot \pi^2}{l^2} - \alpha^2 + \beta^2\right) \geq 0. \tag{43}$$

With (43) being satisfied, it can be checked readily that the vector-valued function $u(t, x)$ defined by (38), (39), and (40) is a solution for (11), (12), (13), and (14). The frequency of this solution is

$$v = \frac{|k|}{2\pi} = \sqrt{\frac{\mu}{\rho}} \cdot \sqrt{\left(\frac{n^2 \cdot \pi^2}{4 \cdot \pi^2 \cdot l^2} - \frac{\alpha^2}{4 \cdot \pi^2} + \frac{\beta^2}{4 \cdot \pi^2}\right)}. \tag{44}$$

● **Case IV**: $f = \sinh$ or $f = \cosh$ with $g = \sinh$. In this case the sets $(\gamma_1, \gamma_2, \gamma_3)$ and $(n, \alpha, \beta, k)$ of parameters must respectively satisfy

$$\gamma_1 \cdot \frac{n \cdot \pi}{l} - \gamma_2 \cdot \alpha - \gamma_3 \cdot \beta = 0 \quad \text{and} \quad k^2 = \frac{\mu}{\rho} \cdot \left(\frac{n^2 \cdot \pi^2}{l^2} - \alpha^2 - \beta^2\right) \geq 0. \tag{45}$$

With (45) being satisfied, it can be checked readily that the vector-valued function $u(t, x)$ defined by (38), (39), and (40) is a solution for (11), (12), (13), and (14). The frequency of this solution is

$$v = \frac{|k|}{2\pi} = \sqrt{\frac{\mu}{\rho}} \cdot \sqrt{\left(\frac{n^2 \cdot \pi^2}{4 \cdot \pi^2 \cdot l^2} - \frac{\alpha^2}{4 \cdot \pi^2} - \frac{\beta^2}{4 \cdot \pi^2}\right)}. \tag{46}$$

When $|\alpha|$ and $|\beta|$ are relatively small, we have the following estimate for the vibration frequency of $u(t, x)$:

$$v \approx \sqrt{\frac{\mu}{\rho}} \cdot \frac{|n|}{2l} \tag{47}$$

for all the cases discussed above.

**Remark 2.** When the requirement $k^2 \geq 0$ is not satisfied in any of the last three cases,

$$k^2 = \frac{\mu}{\rho} \cdot \left(\frac{n^2 \cdot \pi^2}{l^2} \pm \alpha^2 \pm \beta^2\right) \prec 0 \tag{48}$$

so that $k$ is *purely imaginary*, the displacement function $\boldsymbol{u}(t,\boldsymbol{x})$ defined by (38), (39), and (40) is still a solution for (11), (12), (13), and (14). However the corresponding displacement function $\boldsymbol{u}(t,\boldsymbol{x})$ would grow exponentially and could lead to *failure* of rubber wiper.

## 2.4. Physical significance of some parameters of solutions

It has been explained in Remarks 1 and 2 that unfavorable solutions of Class I or Class II might lead to malfunctioning of rubber wiper on windshield.

Now we discuss the physical significance of the parameters $\alpha$ and $\beta$ appearing in the construction of soultions of Class I defined by (15), (16), (17), and (18).

Assume that $\boldsymbol{u}(t,\boldsymbol{x}) = (u_1(t,\boldsymbol{x}), u_2(t,\boldsymbol{x}), u_3(t,\boldsymbol{x}))$ is a *displacement* function of Class I defined by (15), (16), (17), and (18). It can be shown, through calculation using the Elasticity Mechanics (Frankel, 2011; Chaichian et al., 2012), that the $(x_2, x_3)$ *shear stress* of this solution is

$$\alpha \cdot \beta \cdot \sin(\varphi + k \cdot t) \cdot \sin\left(\frac{n \cdot \pi \cdot x_1}{l}\right) \cdot \cos(a + \alpha \cdot x_2) \cdot \cos(\beta \cdot x_3). \tag{49}$$

It can be observed readily that, when $|\alpha|$ and $|\beta|$ are large, the *variation* of the distribution of $(x_2, x_3)$ *shear stress* on the wiper lip becomes large. This means that the *friction force* on the wiper lip becomes large, if the pressure on wiper, the drag force, and the support force from the windshield remain unchanged.

Similarly, it can be shown, through calculation using the Elasticity Mechanics, that the $(x_1, x_3)$ *shear stress* of this solution is

$$\beta \cdot \frac{n \cdot \pi}{l} \cdot \sin(\varphi + k \cdot t) \cdot \cos\left(\frac{n \cdot \pi \cdot x_1}{l}\right) \cdot \sin(a + \alpha \cdot x_2) \cdot \cos(\beta \cdot x_3). \tag{50}$$

Besides, the $(x_1, x_2)$ *shear stress* of this solution is

$$\alpha \cdot \frac{n \cdot \pi}{l} \cdot \sin(\varphi + k \cdot t) \cdot \cos\left(\frac{n \cdot \pi \cdot x_1}{l}\right) \cdot \cos(a + \alpha \cdot x_2) \cdot \sin(\beta \cdot x_3). \tag{51}$$

It can be observed readily that, when $|\alpha|$ and $|\beta|$ are large, the *friction force* on the wiper lip becomes more *uneven*, if the pressure on wiper, the drag force, and the support force from the windshield remain unchanged.

Thus we conclude that, when the *friction force* acting on the wiper lip is not highly irregular, the vibration frequencies of solutions of Class I, discussed in Subsection 2.2, would locate around

$$\sqrt{\frac{\lambda + 2\mu}{\rho}} \cdot \frac{n}{2l}. \tag{52}$$

Similarly we expect that the vibration frequencies of solutions of Class II, discussed in Subsection 2.3, would locate around

$$\sqrt{\frac{\mu}{\rho}} \cdot \frac{n}{2l}. \tag{53}$$

## 3. Comparison of model predictions with experimental data

The material constants of our rubber wiper are listed as follows:
$$\rho = 10^3 \ Kg/m^3, \quad E = 5 \cdot 10^6 \ \text{Pa}, \ \text{and} \ \sigma = 0.49. \tag{54}$$

The length of our rubber wiper is $24" \doteq 0.61\text{m}$. Simple calculation of (3) shows that

$$\frac{1}{2l} \cdot \sqrt{\frac{\lambda + 2\mu}{\rho}} \ \text{Hz} \approx 240 \ \text{Hz} \quad \text{and} \quad \frac{1}{2l} \cdot \sqrt{\frac{\mu}{\rho}} \ \text{Hz} \approx 33.6 \ \text{Hz}. \tag{55}$$

Thus the vibration frequencies of Class I should locate around $n*240 \ \text{Hz} = 240\text{Hz}$, 480Hz, 720Hz, 960Hz, 1200Hz, 1440Hz, 1680Hz, 1920Hz, 2160Hz, 2400Hz, etc. Moreover, the vibration frequencies of Class II should locate around $n*33.6 \ \text{Hz}$.

Our experiments are made for rubber wiper on slightly wet windshield. Since the dynamics of rubber wiper around reversal, where the wiper reverses its moving direction, is much complicated, we only compare our model predictions with the experimental data of the vibration frequencies of rubber wiper away from reversal.

Assume that the wiper starts a "side-to-side cycle" at time $T_1$. Assume that the wiper reverses its moving direction at time $T_2$. Assume that the wiper comes back to its starting position at time $T_3$, to start another cycle. See Figure 4.

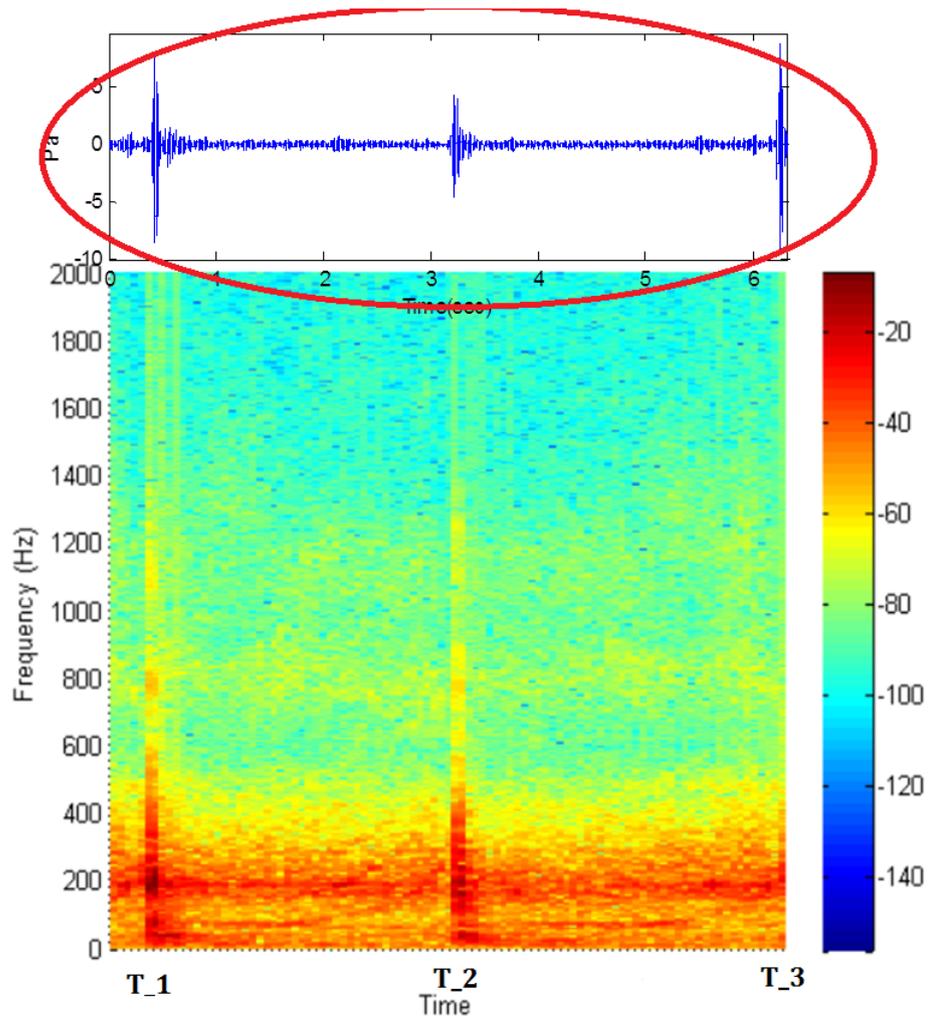

**Figure 4.** Frequencies recorded depending on the time variable.

Let $A = T_2 - T_1$. The FFT (Fast Fourier Transform) diagram of the fequencies recorded between $T_1 + (0.25) \cdot A$ and $T_1 + (0.75) \cdot A$ is shown in Figure 5. In this FFT diagram Figure 5, magnitude of FFT for frequencies below 25,600Hz can be trusted.

It can be observed in Figure 5 that the predicted peaks of frequencies, 240Hz, 480Hz, 720Hz, 960Hz, 1200Hz, 1440Hz, 1680Hz, 1920Hz, 2160Hz, 2400Hz of Class I, appear manifestly.

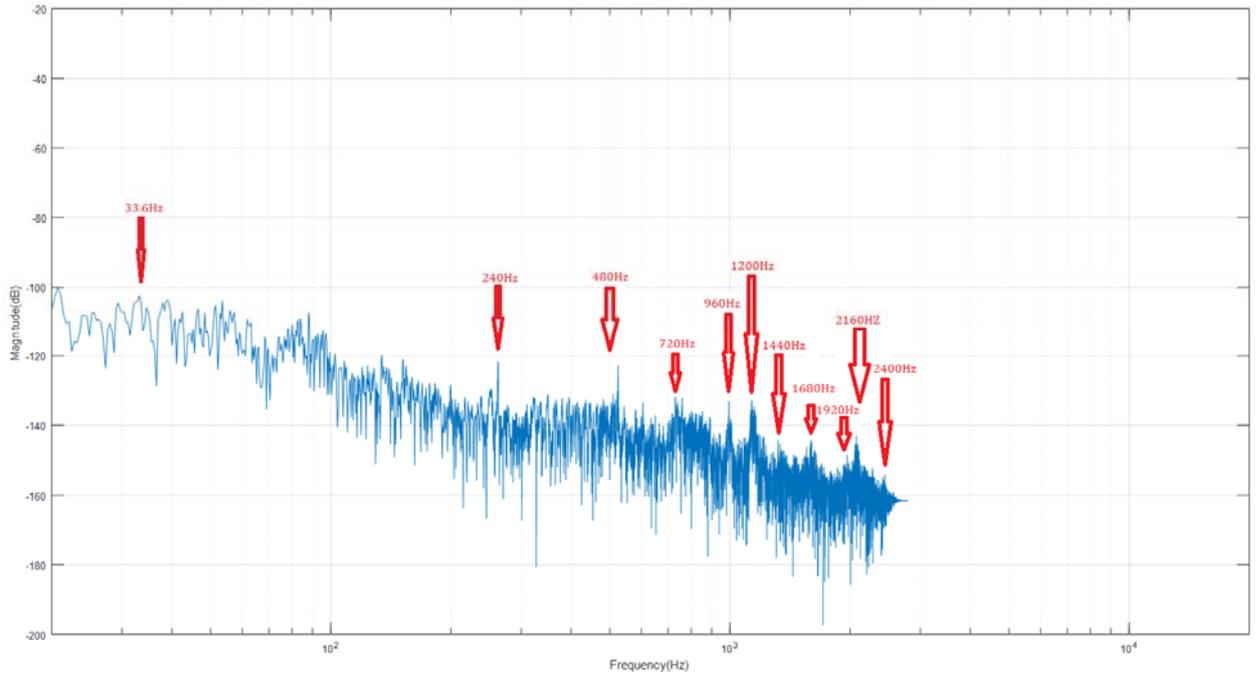

**Figure 5.** Vibration frequencies of wiper before reversal.

Figure 6 shows the FFT diagram of background /machine/white noise recorded between $T_1 + (0.25) \cdot A$ and $T_1 + (0.75) \cdot A$. In Figure 6, magnitude of FFT for frequencies below 25,600Hz can be trusted.

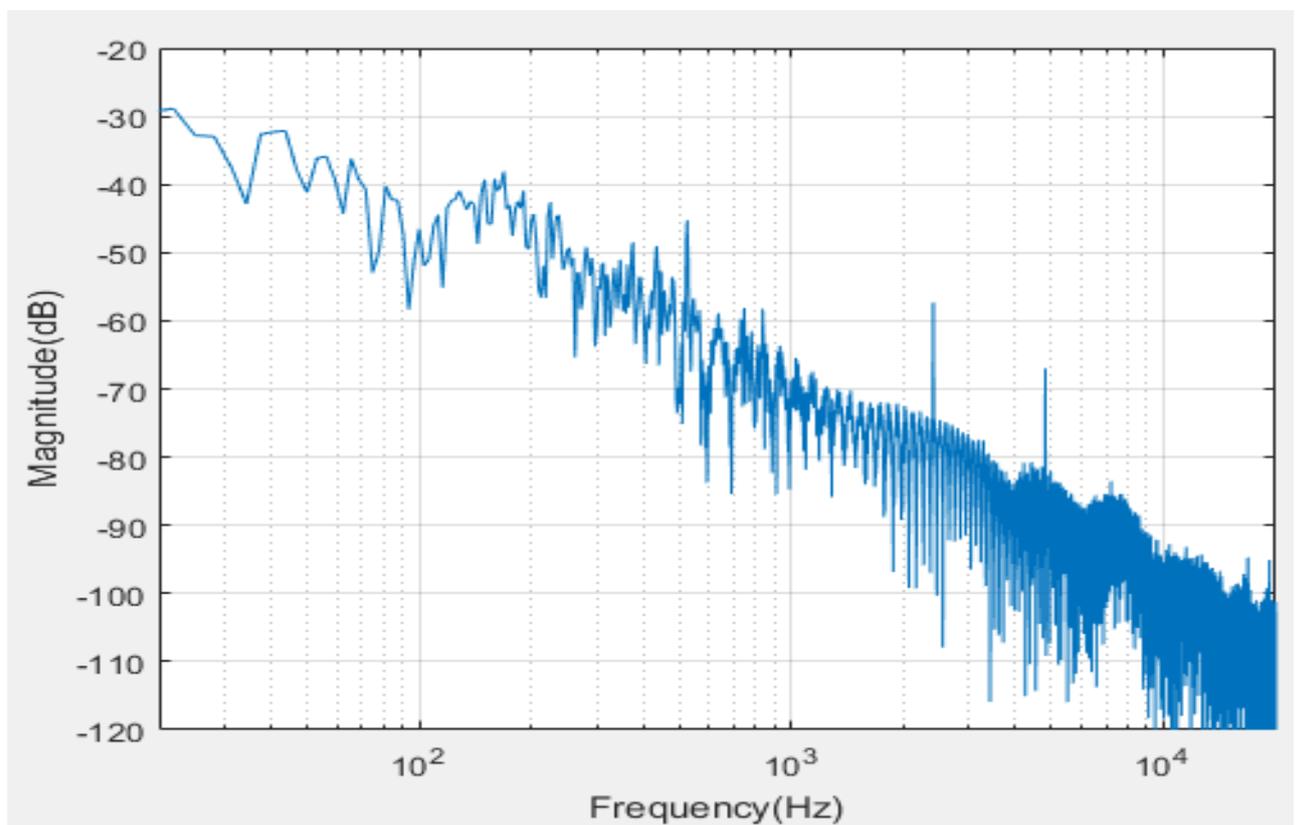

**Figure 6.** Frequencies of "background/machine/white noise" before reversal.

Let $B = T_3 - T_2$. The FFT diagram of the frequencies recorded between $T_2 +(0.25)\cdot B$ and $T_2 +(0.75)\cdot B$ is shown in Figure 7. In this FFT diagram Figure 7, magnitude of FFT for frequencies below 25,600Hz can be trusted.

In Figure 7, the predicted peaks of frequencies, 33.6Hz, 240Hz, 720Hz, 1440Hz, 1680Hz, appear manifestly. Here the highest peak appears at the frequency 33.6Hz of Class II.

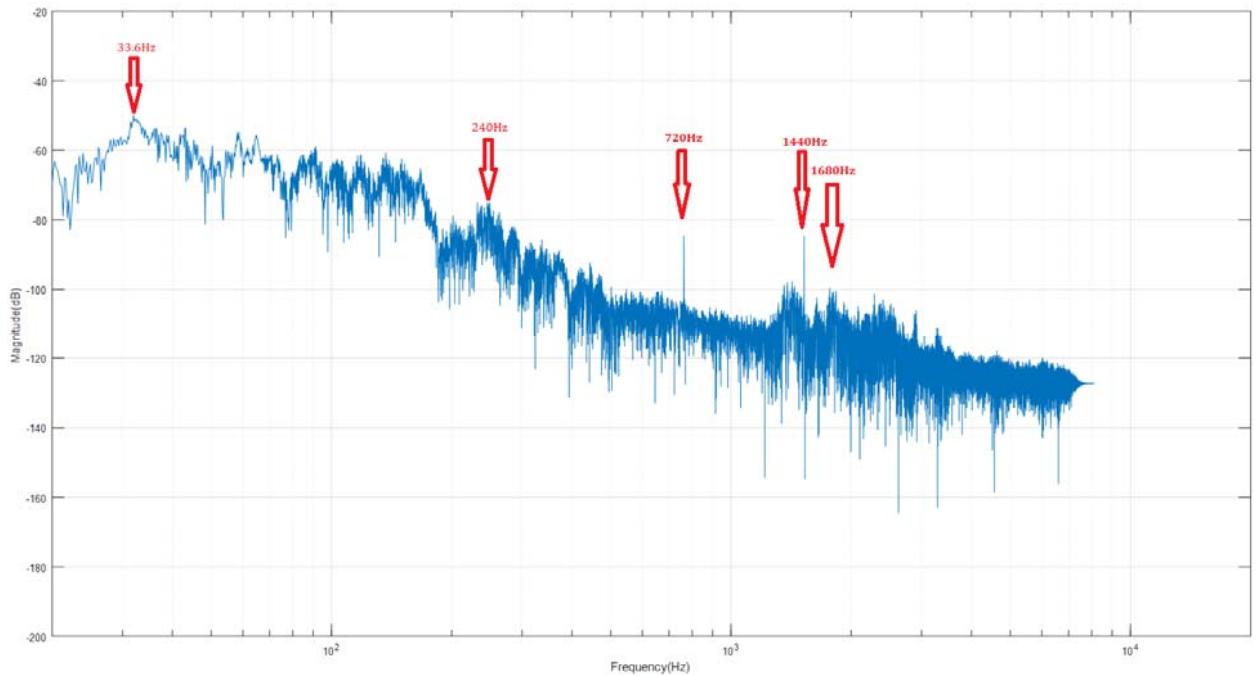

**Figure 7.** Vibration frequencies of wiper after reversal.

Figure 8 is the FFT diagram of "background/machine/white noise" recorded between $T_2 +(0.25)\cdot B$ and $T_2 +(0.75)\cdot B$. In Figure 8, magnitude of FFT for frequencies below 25,600Hz can be trusted.

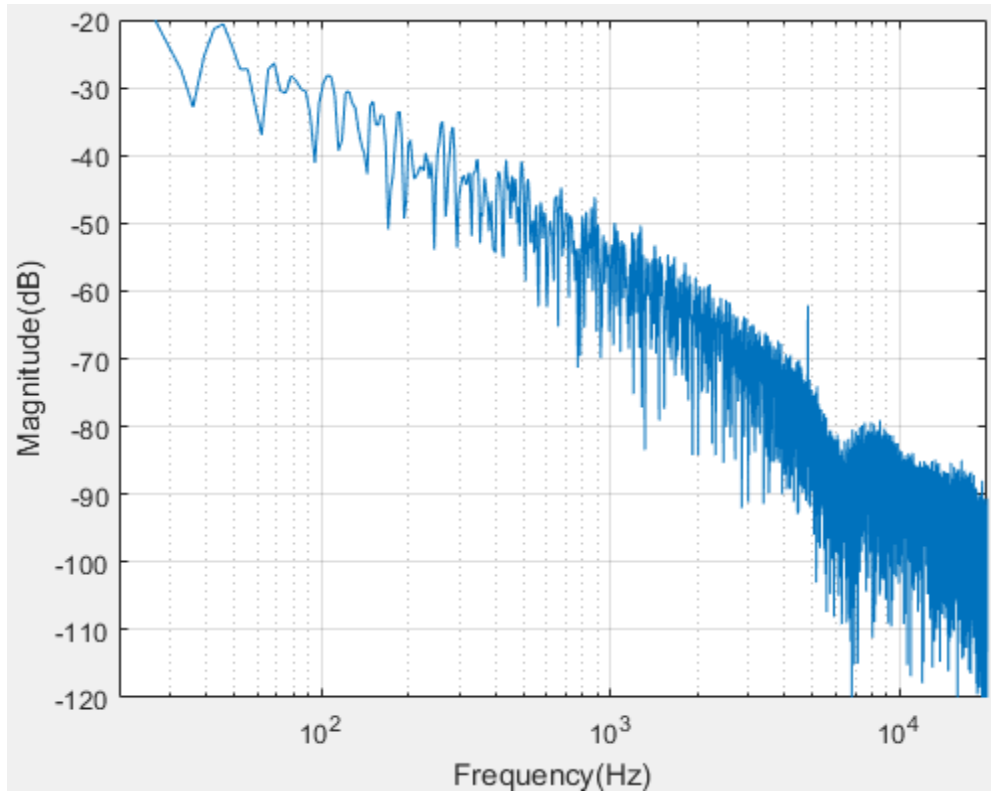

**Figure 8.** Frequencies of "background/machine/white noise" after reversal.

## 4. Conclusion and discussions

Comparison of our model predictions with experimental data shows the accuracy of our mathematical formulae for the vibration frequencies of rubber wiper on windshield. As is explained in Subsection 2.4, our model predictions will be in excellent agreement with experimental data, when the *friction force* acting on the wiper lip is not highly irregular. It turns out that our mathematical formulae can be used to test the quality of wiper design on the evenness of friction force. Further implications of our analysis will appear elsewhere.

**Acknowledgements**
We are very grateful to Professors Chien-Hsiung Tsai, Hsi-Wei Shih, Min-Hung Chen, Yu-Chen Shu, and Chyuan-Yow Tseng for many stimulating discussions.

## Declaration of Conflicting Interests
The author(s) declared no potential conflicts of interest with respect to the research, authorship, and/or publication of this article.

## Funding
The author(s) received no financial support for the research, authorship, and/or publication of this article.

**References**

Okura S, Sekiguchi T and Oya T (2000) Dynamic analysis of blade reversal behavior in a windshield wiper system. *Sea Technical Paper Series*: 2000-01-0127.

Goto S, Takahashi H and Oya T (2001) Clarification of the mechanism of wiper blade rubber squeal noise generation. *JSAE REVIEW* 22: 57-62.

Grenouillat R and Leblanc C (2002) Simulation of chatter vibrations for wiper systems. *Sea Technical Paper Series*: 2002–01–1239.

Stein GJ, Zahoransky R and Múčka P (2008) On dry friction modelling and simulation in kinematically excited oscillatory systems. *Journal of Sound and Vibration* 311: 74–96.

Lancioni G, Lenci S and Galvanetto U (2009) Non-linear dynamics of a mechanical system with a frictional unilateral constraint. *International Journal of Non-Linear Mechanics* 44: 658-674.

Sugita M, Yabuno H and Yanagisawa D (2012) Bifurcation phenomena of the reversal behavior of an automobile wiper blade. *Nonlinear Dynamics* 69: 1111-1123.

Min D, Jeong S, Yoo H, et al. (2014) Experimental investigation of vehicle wiperblade's squeal noise generation due to windscreen waviness. *Tribology International* 80: 191-197.

Reddyhoff T, Dober O, Rouzic JL, et al. (2015) Friction induced vibration in windscreen wiper contacts. *Journal of Vibration and Acoustics* 137: 1–7.

Lancioni G, Lenci S and Galvanetto U (2016) Dynamics of windscreen wiperblades: squealnoise, reversal noise and chattering. *International Journal of Non-Linear Mechanics* 80: 132-143.

Unno M, Shibata A, Yabuno, H, et al. (2017) Analysis of the behavior of a wiper blade around the reversal in consideration of dynamic and static friction. *Journal of Sound and Vibration* 393: 76-91.

Hong Y-J (1998) Ruled Manifolds with Constant Hermitian Scalar Curvature. *Mathematical Research Letters* 5: 657-673.

Hong Y-J (1999) Harmonic Maps into the Moduli Spaces of Flat Connections. Annals of Global Analysis and Geometry 17: 441-473.

Hong Y-J (1999) Constant Hermitian scalar curvature equations on ruled manifolds. *Journal of Differential Geometry* 53: 465-516.

Hong Y-J (2002) Gauge-Fixing Constant Scalar Curvature Equations on Ruled Manifolds and the Futaki Invariants. *Journal of Differential Geometry* 60: 389-453.

Hong Y-J (2008) Stability and existence of critical Kaehler metrics on ruled manifolds. *Journal of the Mathematical Society of Japan* 60: 265-290.

Hong Y-J and Chen T-J (2020) Geometric aspects of the Lame equation. *In Preparation*.

Frankel T (2011) *The Geometry of Physics: An Introduction*. Cambridge University Press.

Chaichian M, Merches I and Tureanu A (2012) *Mechanics: An intensive course*. Springer.

Reissner E (1985) Reflections on the theory of elastic plates. *Applied Mechanics Reviews* 38: 1453-1464.

Hetnarski RB and Ignaczak J (2004) *Mathematical Theory of Elasticity*. Taylor & Francis.

Serov V (2017) *Fourier Series, Fourier Transform and Their Applications to Mathematical Physics*. Springer.

Strauss WA (2008) *Partial Differential Equations: An Introduction*. Wiley.

Thompson PA, Robbins MO (1990) Origin of stick-slip motion in boundary lubrication. *Science* 250:792-794.

Persson BNJ (1994) Theory of friction: the role of elasticity in boundary lubrication. *Physical Review B* 50: 4771-4787.

Yang C and Persson BNJ (2008) Molecular dynamics study of contact mechanics: contact area and interfacial separation from small to full contact. *Physical Review Letters* 100: 024303.

Gilbarg D and Trudinger NS (2001) *Elliptic Partial Differential Equations of Second Order*. Springer.

Jost J (2013) *Partial Differential Equations*. Springer.